\documentclass[prd,preprintnumbers,amsmath,amssymb,showpacs]{revtex4}

\usepackage{graphicx}
\usepackage{epsfig}
\usepackage{psfrag}
\usepackage{rotating}
\usepackage{dcolumn}
\usepackage{bm}

\newcommand{\lsim}{\raisebox{-0.13cm}{~\shortstack{$<$ \\[-0.07cm] $\sim$}}~} 
\newcommand{\gsim}{\raisebox{-0.13cm}{~\shortstack{$>$ \\[-0.07cm] $\sim$}}~} 
\newcommand{\ra}{\rightarrow}

\newcommand{\beq}{\begin{eqnarray}} 
\newcommand{\eeq}{\end{eqnarray}}

\begin{document}

\preprint{LPT-ORSAY-07-52}

\title{Higgs production at the LHC in warped extra--dimensional models}
\author{Abdelhak Djouadi}
\affiliation{Laboratoire de Physique Th\'eorique, U. Paris--Sud and CNRS, 
F--91405 Orsay, France.}
\author{Gr\'egory Moreau}
\affiliation{Laboratoire de Physique Th\'eorique, U. Paris--Sud and CNRS, 
F--91405 Orsay, France.}

\begin{abstract}   

The extra--dimensional model in which the bulk geometry is a slice of  anti--de
Sitter space is a particularly attractive extension of the Standard Model  as it
allows to address the gauge hierarchy problem, as well as the mass hierarchy
prevailing among fermions.  However, to allow for the masses of the
Kaluza--Klein excitations of the known particles to be near the Terascale
without conflicting with the high--precision electroweak data, one needs to
promote the gauge symmetry to a left--right  structure ${\rm SU(2)_L \! \times
\! SU(2)_R \! \times\! U(1)}$ which incorporates a new quark $b'$, the ${\rm
SU(2)_R}$ doublet partner of the heavy top quark. We show  that this new quark
will contribute to the main production process of Higgs bosons at the LHC: the
gluon--gluon fusion mechanism which proceeds through heavy quark triangular
loops.  In most of the parameter  space in which the measured values of the
heavy $t,b$ quark masses are reproduced,  the $gg \to\;$Higgs production cross
section is significantly altered, even if the $b'$ quark is too heavy to be
directly accessible, $m_{b'} \gsim 1$ TeV. Finally, we briefly discuss the new
Higgs production and decay channels involving the $b'$ quark. 

\end{abstract}   
\maketitle

\large

The warped extra--dimensional scenario proposed by Randall and Sundrum (RS) 
\cite{RS}  is a  particularly attractive extension of the Standard Model (SM).
In addition to its original motivation, to provide a solution to the gauge
hierarchy  problem, this scenario  turned out to represent a suitable framework 
to address several important  phenomenological issues. For instance, the
version of the model with bulk matter provides new weakly interacting candidates
for dark matter in the universe  \cite{LZP2}, allows for the  unification of the
gauge couplings at high--energies \cite{UNI-RS}  and proposes a new
interpretation of SM fermion mass hierarchies  based on specific localizations
of the fermion  wave functions along the warped extra dimension  \cite{RSloc}. 

Indeed, if the fermions are placed differently along the extra dimension,
hierarchical patterns among the effective four--dimensional  Yukawa couplings
are generated as a result of their various wave function overlapping with the
Higgs boson, which remains  confined on the so--called TeV--brane for its mass
to be protected. A parameter denoted $c_f$ quantifies the five--dimensional
mass  affected to each fermion representation, $\pm c_f k$ where $1/k$ is the
anti--de Sitter $({\rm AdS_5})$ curvature radius,  and fixes the fermion
localization with respect to the TeV--brane. As the parameter $c_f$ decreases,
the zero--mode fermions become closer to the TeV--brane and acquire  a larger
mass.  Remarkably, this geometrical mechanism of mass generation is possible 
for values $\vert c_f \vert \sim 1$, i.e. for fundamental mass parameters of the
order of the unique scale of the theory: the reduced Planck mass scale
$M_{P}\sim k$.

If this extra--dimensional model is to solve the gauge hierarchy problem, the
masses of the first Kaluza--Klein (KK) excitations of the SM gauge bosons must
be in the vicinity of the TeV scale but such low masses lead to  unacceptably
large contributions \cite{Burdman} to the precisely measured  electroweak
observables \cite{PDG}. Nevertheless, it was shown \cite{ADMS} that if the SM
gauge symmetry is enhanced to the left--right  custodial structure ${\rm
SU(2)_L\! \times\! SU(2)_R\! \times\! U(1)}$ (or, alternatively, to the O(4)
custodial structure \cite{AFB-Rold}), the high--precision data can be fitted
while keeping  the KK gauge boson masses down to an acceptable value, $M_{\rm
KK}  \sim 3$ TeV, for light fermion localizations  $c_{\rm light} \gtrsim 0.5$;
the heavy  $b$ and $t$ quarks must be treated separately \cite{Carena}. In
fact, since third generation fermions interact more strongly with the KK gauge
bosons, that mix with the $Z$--boson, one can even solve the anomaly observed
in  the forward--backward $b$--quark asymmetry measured at LEP
\cite{AFB-Rold,RSAFB}, the only (high--energy)  observable that deviates
significantly from the SM prediction.  

In this extension of the SM group, the right--handed fermions are  promoted to
${\rm SU(2)_R}$ isodoublets. A new quark $b'_R$,  the ${\rm SU(2)_R}$ 
partner of the right--handed top quark $t_R$, is present  and should be
typically  much lighter than the other ${\rm SU(2)_R}$  partners and all KK
excitations of the SM fermions \cite{LZP2}, the latter ones being by 
construction heavier than the KK gauge bosons.  Indeed, as the ${\rm SU(2)_R}$
symmetry is broken by boundary conditions \cite{ADMS},  the $b'_R$ quark has  
Dirichlet boundary conditions on the Planck--brane and Neumann ones on the
TeV--brane \cite{Csaki}, noted $(-+)$,  so that it has no zero--mode,   in
contrast to the SM fermions which have $(++)$ boundary conditions. Moreover, the
mass of the first  KK excitation of the $b'_R$ should be relatively low as it 
is controlled by the same $c_t$ parameter as the top quark $t_R$ which must be
sufficiently small in order to generate a large $m_t$ value. 

\begin{figure}[!h]
\begin{center}
\includegraphics[width=0.8\linewidth,bb=73 680 600 755]{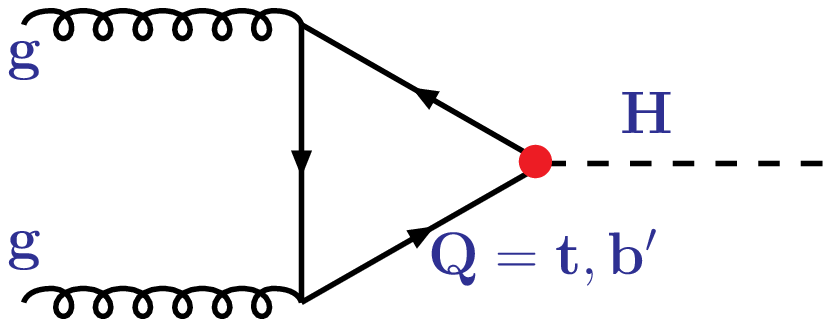} 
\end{center}
\caption[]{Feynman diagram for the $gg\!\to\!H$ production process.}
\label{fig:ggH} 
\end{figure}

In this scenario, the presence of the heavy new quark, although its Higgs
coupling is not proportional to the KK mass,  could significantly alter the
phenomenology of the Higgs particle which, otherwise, is expected to have
similar properties as the SM $H$ boson \cite{Higgsrev}.  As a matter of fact,
the $b'$ state will mix with the $b$ quark and will slightly modify the $Hb\bar
b$ Yukawa coupling which controls the decay branching ratios of the Higgs boson
if its mass is in the range $M_H\lsim 140$ GeV.  A second and more drastic
consequence of the presence of a relatively light $b'$ state is that it will 
contribute to the main production channel for a SM--like Higgs boson at the
Large Hadron Collider (LHC): the gluon--gluon fusion mechanism $gg\to H$. This
process proceeds through a triangular loop built up by heavy quarks,
Fig.~\ref{fig:ggH}, and in the SM only the contribution of the top quark is
significant as a result of the much larger Yukawa coupling (the $b$--quark
contribution is smaller than 10\% even for low Higgs masses). 

We will show in this paper that the $b'$ contribution to the $Hgg$ loop
amplitude can be significant for model parameters that allow to reproduce the
measured $m_b$ and $m_t$ values. The Higgs
production cross section in the $gg \to H$ process can be enhanced by a factor
of four  compared to its SM value.

This is in contrast to the other fermion KK excitations which decouple from the
$Hgg$ amplitude as they are expected to be much heavier \cite{BL}. 
Note, however, that within the universal extra--dimensional
scenario where the Higgs boson propagates in the bulk,  the KK fermion modes can
be light enough to induce a sizable  modification of the  $gg \to H$ production
rate \cite{UEDpre}.

To calculate the $Hgg$ amplitude, one needs first to derive the $t,b'$ 
couplings to the Higgs boson. The Yukawa terms take an invariant form under
the custodial symmetry, as the SM ${\rm SU(2)_L}$ quarks are singlets under
${\rm SU(2)_R}$ whereas the $H$ field is embedded into a bidoublet.  The whole
$b$--quark mass matrix needs to be studied as the Yukawa couplings induce
mixings with the KK excitations; the $t$--quark mass matrix has a similar
structure. For simplicity, we describe only the largely dominant third
family contribution and consider only the first KK excitations that we denote
with the exponent in brackets $n\!=\!1,2,\dots$ which labels the KK--level. In
the field basis  $\Psi_L^t \equiv (b^{(0)}_L, b^{(1)}_L, b^{c(1)}_L,
b^{\prime(1)}_L, b^{\prime(2)}_L)^t$,  $\Psi_R^t \equiv (b^{c(0)}_R, b^{(1)}_R,
b^{c(1)}_R, b^{\prime(1)}_R, b^{\prime(2)}_R)^t$ where we introduce the charge
conjugated fields (indicated by the superscript $c$)  in order to use only
left--handed SM fields, the effective four--dimensional bottom quark  mass terms
are of Dirac type,       ${\cal L}_{\rm m}\!=\!\bar \Psi_L  {\cal M}_b
\Psi_R\!+ \!{\rm h.c.}$. After electroweak symmetry breaking, one obtains for
the $5\times 5$ mass matrix  ${\cal M}_b$, in terms of the $c_Q$ and $c_b$
parameters associated, respectively, to the SM doublet
$Q_L^t\!=\!(t_L,b_L)^t$ and singlet $b^c_R$:
\begin{eqnarray} 
{\cal M}_b = \hspace*{0.5cm} 
\left(  
\begin{array}{ccccc}  \tilde v f^{\star(0)}_{c_Q} f^{(0)}_{c_b} & 0 & \tilde v
f^{\star(0)}_{c_Q} f^{(1)}_{c_b} &  \tilde v f^{\star(0)}_{c_Q} g^{(1)}_{c_t} &
\tilde v f^{\star(0)}_{c_Q} g^{(2)}_{c_t} \\  \tilde v f^{\star(1)}_{c_Q}
f^{(0)}_{c_b} & m^{(1)}_{c_Q} & \tilde v f^{\star(1)}_{c_Q} f^{(1)}_{c_b} &   
\tilde v f^{\star(1)}_{c_Q} g^{(1)}_{c_t} & \tilde v f^{\star(1)}_{c_Q}
g^{(2)}_{c_t} \\  0 & 0 & m^{(1)}_{c_b} & 0 & 0 \\  0 & 0 & 0 &
m^{\prime(1)}_{c_t} & 0 \\  0 & 0 & 0 & 0 & m^{\prime(2)}_{c_t}   
\end{array} \right)  
\label{eq:massmat}  
\end{eqnarray}   
with $\tilde v = - \lambda_5 v / \sqrt{2} \pi R$, $\lambda_5$ being the 
five--dimen\-sional dimensionful Yukawa coupling constant, ta\-ken equal to
$k^{-1}$  as usual to avoid the introduction of an additional scale, $v \simeq
246$ GeV being the Higgs vacuum expectation value and $R$ the compactification
radius. $f^{(n)}_{c}/\sqrt{\pi R}$ and $g^{(n)}_{c}/\sqrt{\pi R}$   stand for
the wave functions of the $n$--th KK mode of a field  characterized by the $c$
parameter and, respectively, $(++)$ and $(-+)$  boundary conditions;   all wave
function values are taken at the position of the TeV--brane,   $x_5\!=\!\pi R$.
Finally, $m^{(n)}_{c}$ and $m^{\prime(n)}_{c}$ are the $n$--th   KK masses for,
respectively, the $(++)$ and  $(-+)$ fields; $m^{\prime(1)}_{c}$  decreases with
the parameter $c$ and can reach particularly small values  \cite{LZP2}. The
zeroes in the matrix eq.~(\ref{eq:massmat}) originate from the fact that the 
fields $b^{(1)}_R$, $b^{c(1)}_L$ and $b^{\prime(n)}_L$ (with $n=1,2$) have
Dirichlet boundary conditions on the TeV--brane and, thus, do not couple to  the
Higgs boson. 

The matrix (\ref{eq:massmat}) is diagonalized by unitary matrices $U_{L/R}$,
through the transformation  $\Psi'_{L/R}\!=\!U_{L/R}\Psi_{L/R}$ :   
\beq
U_{L} {\cal M}_b U_{R}^\dagger \ = \ {\rm diag}~(m_{b_1},m_{b_2},\dots)
\eeq
where $m_{b_1}$ corresponds to the measured value of the bottom quark mass  (we
take the unitary matrices  such that $m_{b_1}\!<\!m_{b_2}\!<\!\dots$). The 
$\Psi'_{L/R}$ components are the mass eigenstates, namely  $\Psi_{L/R}^{\prime
t} \equiv (b_{1 L/R},b_{2 L/R},\dots)^t$, where generally  $b_{1 L/R}$ is 
mainly composed of the bottom quark zero--mode and $b_{2 L/R}$ of the
$b^{\prime (1)}_{L/R}$. Then, the Higgs couplings are  given by the interaction
Lagrangian 
\beq
{\cal L}_{\rm int} =  \frac{H}{v} \  \bar \Psi'_L {\cal M}''_b \Psi'_R \ + \ 
{\rm h.c.}
\eeq
where ${\cal M}''_b = U_{L} {\cal M}'_b U_{R}^\dagger$ and ${\cal M}'_b$ is the 
matrix of eq.~(\ref{eq:massmat}) but with the KK masses set to zero.  Hence, the
Higgs coupling to an eigenstate $b_{iL/R}$ is given by ${\cal M}''_{bii}/v$ in
contrast to the usual $m_b/v$ value in the SM.  Besides, the orthonormality
condition for the fermion wave functions,  together with the flatness of the 
gluon zero--mode wave function, lead to a gluon coupling with $b_{i L/R}$
states which is identical to the SM quark--gluon coupling. 

In order to obtain the value for each effective quark coupling to the Higgs
boson, we have diagonalized numerically the mass matrix (1);   to reach a good
degree of convergence and an accurate determination of the quark masses and
couplings, we have included states up to $b_5$ and $t_5$ in the sum over KK
modes.

We are now in a position to discuss the new state contributions to the
$gg\!\to\!H$ production rate at the LHC. To lowest order, the partonic cross 
section reads \cite{Hgg-LO}
\beq 
\sigma_H = \frac{G_F\alpha_{s}^{2} M_H^2}{288 \sqrt{2}\pi} \, \left| \, 
\sum_{Q} A_{1/2}^H (\tau_{Q}) \, \right|^{2} \, \delta (\hat s -M_H^2) 
\eeq
where $\hat{s}$ is the $gg$ invariant energy squared.  The form factor
$A_{1/2}^H (\tau_Q)$ with $\tau_Q = M_H^2/4m_Q^2$ is normalized such that for
$m_Q\!\gg\!M_H$, it reaches unity while it approaches zero in the  chiral limit
$m_Q \ra 0$. In fact, the approximation $A_{1/2}^H \simeq 1$ is  very good for
Higgs masses below the heavy quark threshold, $M_H \lsim 2m_Q$ \cite{Hgg-HO}, 
Since high--precision data suggest that the Higgs boson is relatively light, 
$M_H\lsim 200$ GeV \cite{PDG}, and the new quarks are expected to be heavier than the top 
quark (otherwise they would have probably been observed at the Tevatron \cite{PDG})  
this approximation holds and will be adopted here.  

It is convenient to consider the ratio ${\cal R}=\sigma_H^{\rm RS}/\sigma_H^{\rm
SM}$ of the $gg \to H$ production cross sections in the RS and SM models which,
including the contributions  of the first KK excitations 
reads 
\begin{eqnarray}
{\cal R} \equiv  \frac{ \sigma^{\rm RS}_H }{\sigma^{\rm SM}_H } \simeq
\bigg \vert 
\sum_{i=1}^5 \frac{{\cal M}''_{tii}}{m_{t_i}} +
\sum_{j=2}^5 \frac{{\cal M}''_{bjj}}{m_{b_j}}
\bigg \vert^2, 
\label{eq:ratio}
\end{eqnarray}
if we neglect the relatively small $b$--quark contribution and use the
approximation $A_{1/2}^H \simeq 1$ discussed above. Note that the higher order
QCD corrections, which are known to be rather large \cite{Hgg-HO}, are 
essentially the same for all quark species and, thus, drop in the ratio ${\cal
R}$. 

\begin{figure}[!h] 
\begin{center} 
\psfrag{cQL}[c][c][1]{{\large $\mathbf{c_Q}$}} 
\psfrag{ctR}[c][c][1]{{\large $\mathbf{c_t}$}} 
\epsfig{file=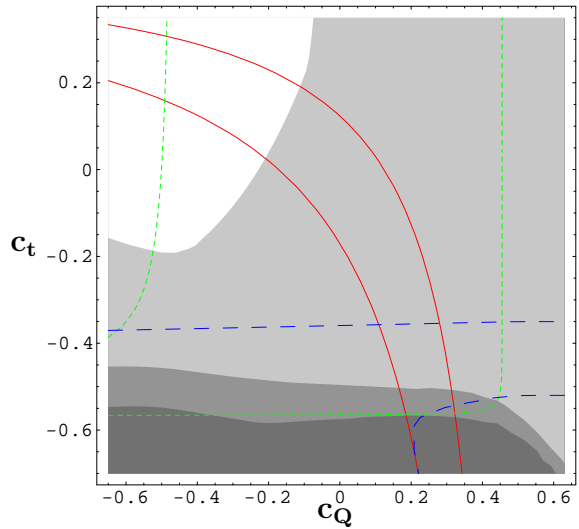,width=7.6cm} 
\end{center} 
\vspace*{-3mm}
\caption{Values of the ratio ${\cal R}=\sigma_H^{\rm RS}/\sigma_H^{\rm SM}$ in
the plane $[c_t,c_Q]$ for a fixed value $c_b\!=\!0.6$. 
The filled regions correspond from  white to
darkest grey to, respectively, the intervals ${\cal R} \in [1,1.2]$, $[1.2,2]$,
$[2,4]$ and ${\cal R}>4$. The
green dotted--lines are the contour levels for $m_{b}\!=1\!$--4.5 GeV, the red
solid--lines for $m_{t}\!=\!150$--200 GeV (both from right to left), while the 
blue dashed--lines are associated to $m_{b_2}= 200$--1000 GeV (from down to up).
We have set $kR= 10.11$ and $k$ such that $M_{\rm KK}= 3$~TeV and restricted
to  values $\vert c \vert ={\cal O}(1)$.}  \label{fig:constructive} 
\end{figure} 

Figure~\ref{fig:constructive} displays domains of the $[c_Q,c_t]$  parameter space
corresponding to certain values of the ratio  ${\cal R}$ for the choice
$c_b\!=\!0.6$;  also represented are the regions in which the bottom and top
quark masses $m_{b}$ and $m_{t}$  are close to their measured value \cite{Mass}.
These masses typically increase as the associated $c$ parameters decrease, as
a     result of the geometrical mechanism for zero--modes mentioned previously, 
and the change of regime in $m_{b}$  for $c_t\!\lsim\!-0.4$ is due to
$b^{(0)}$--$b^{\prime(1)}$ mixing. In most of the regions consistent with the correct
range for both $m_t$ and $m_b$, the $gg \! \to\! H$ cross section in the RS
model is enhanced with respect to the SM case as a result of a
constructive interference of the $t_1$ and $b_2$ loop contributions
\footnote{Note that these interferences are always constructive as the
possibly negative sign of the effective Yukawa coupling is systematically 
compensated by the same--sign mass term needed to flip the chirality for 
each fermion in the loop.}. In these
areas, the effective $Ht \bar t$  coupling is quasi unaffected  (i.e. at most at
the percent level only) by  mixing effects so that almost no correction to the
SM  amplitude is generated from the top quark exchange. The
deviation of the cross section is almost entirely due to the exchange of the 
$b_2$ state in the loop as its mass is smaller than the other heavy states. In
the regions with realistic $t,b$ masses, $m_{b_2}$ can be as low as  several
hundred GeV (for $c_t \lesssim -0.3$ leading to $m^{\prime(1)}_{c_t} \lsim 1$
TeV) while e.g. $m_{t_2}>3$~TeV.  For $m_{b_2}$ values close to 200 GeV,
the $gg \to H$ production rate at the LHC is enhanced by a factor around 4
compared to the SM case.

For $c_b$ values smaller than  the one used above, $c_b \lsim 0.6$, similar
significant deviations to the $gg \to H$ cross section occur. Choosing a $c_b$
value smaller than $0.5$ and in turn more far from the $c$ values for light
quarks would tend to increase the  flavor changing neutral currents in the third
generation, induced at the tree--level by KK gauge bosons. On the other hand,
for values significantly larger than  $c_b=0.6$, it becomes difficult to
generate a sufficiently large mass for the $b$ quark.

Note also that the exact size of the effect could be different in models where
another custodial symmetry (such as ${\rm O(4)}$ for instance) or different
fermion representations (e.g. a certain freedom in the $b_R$ embedding might
exist \cite{AFB-Rold,RSAFB})  are assumed, as  both can affect the quark mass
matrices via the  modification of Clebsch--Gordan  coefficients or the
introduction of  mixing terms with new ``custodial'' partners (custodians). 
Nevertheless, the general aspect is that, if $t_R$ is not a singlet under 
the necessary custodial symmetry, its custodial partner(s) is (are) expected to be 
relatively light, 
in order to generate a sufficiently large $m_t$, and in turn give rise to 
potentially significant effects in the loop--level Higgs production. 
Here, we have studied the minimal custodial symmetry with the simplest
fermion embedding, but clearly, some interesting extended versions 
of the RS model would deserve the same analysis on the Higgs production
at LHC.

As in the case of top quarks, the Higgs coupling to bottom quarks is almost
unchanged in this scenario (some effects would appear only for very low $c_t$
values). Thus, no significant change is  expected in the main Higgs decay
branching ratios, $H\!\to\! b \bar b$ and $H\!\to\!WW,ZZ$ for $M_H$ respectively
below and above $140$  GeV. However, the gluonic  decay $H\!\to\! gg$ is
modified similarly to the $gg \to H$ cross section as it is generated  by the
same amplitude. Furthermore, the rare decays $H\to \gamma\gamma$ and $\gamma Z$
are also  mediated by heavy particle loops and will  be affected by the  $b_2$
state but, contrary to the $Hgg$ case, the dominant component of the
corresponding amplitudes stems from the $W$--boson loop which is SM--like and
e.g.,  the $H\to \gamma \gamma$ decay branching ratio is  affected  only at the
level of a few percent at most. 

\begin{figure}[!h] 
\begin{center} 
\psfrag{Br}[c][c][1]{{\large $Br(H)$}} 
\psfrag{mh}[c][c][1]{{\large $m_{H}$ (GeV)}}
\psfrag{WW}[c][c][1]{{\normalsize $W^+W^-$}}
\psfrag{ZZ}[c][c][1]{{\normalsize $ZZ$}}
\psfrag{tt}[c][c][1]{{\normalsize $t \bar t$}}
\psfrag{bb}[c][c][1]{{\normalsize $b_2 \bar b_2$}}
\epsfig{file=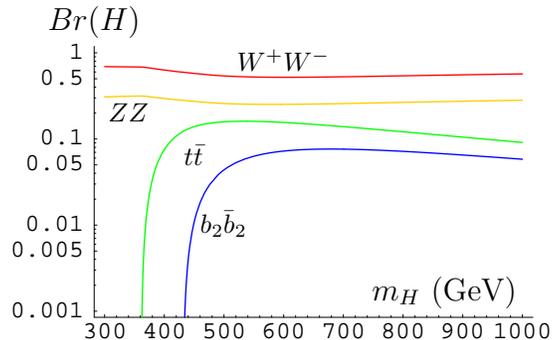,width=7.5cm} 
\end{center} 
\caption{Main branching ratios for the Higgs boson decay channels
[$H \to W^+W^-$; $ZZ$; $t_1 \bar t_1$; $b_2 \bar b_2$] 
as a function of the Higgs mass [in GeV],
for the parameter set: $c_b\!=\!0.6$, $c_t\!=\!-0.55$, $c_Q\!=\!0.24$
which corresponds to $m_{t_1}\!=\!180$ GeV and $m_{b_2}\!=\!215$ GeV.} 
\label{fig:ratio} 
\end{figure} 

We would also like to make the following remark: if the $b_2$ quark mass is
close to the lower value allowed by collider searches, $m_{b_2}\!\sim\!200$ GeV,
and the Higgs boson turns out to be heavy enough (for instance, as a result of
additional corrections to electroweak observables from the light $b'$
\cite{Francois}),  the new decay channel $H\!\to\! b_2 \bar b_2$ could be
kinematically open. This channel could reach a branching ratio quite close to
the $t_1 \bar t_1$ one, as shown in Fig.~(\ref{fig:ratio}). Whether this
scenario is possible and to which extent the Higgs searches might be affected by
these new channels deserve more detailed studies.

In addition, the deviations of the  cross section for associated Higgs
production with top quarks at the LHC, $pp \to Ht\bar  t$, are also expected  to
be small. The other Higgs production channels, vector boson fusion and
associated Higgs production with vector bosons, are of course not affected.
However,  an additional process,  associated  Higgs production with a pair of
$b'$ quarks, $pp \to Hb' \bar b'$, arises. In the regions of parameter space
where the $gg\to H$ rate is enhanced by a factor $\sim 4$, i.e.  for low
$m_{b_2}$ values and large $Hb_2\bar b_2$ couplings,  the cross section for this
new process is similar in size as the one for  $pp \to Ht\bar t$.  
More precisely, for $c_b=0.6$ (the dependence on the $c_b$ value being weak), 
the cross section ratio $\sigma(pp \to Hb_2\bar b_2)/\sigma(pp \to Ht\bar t)$
is equal to the squared ratio of effective Higgs couplings to $b_2$ and $t$ which is
equal to unity if $m_{b_2}=m_t=180$ GeV (which in turn fixes the $c_Q$ and $c_t$ 
parameters as exhibits Fig.~\ref{fig:constructive}), independently of the Higgs mass.

The maximum effects of the new $b'$ state on the Higgs production and
decays occur mainly for $b'$ masses around a few hundred GeV. It might turn out
that, after a precise fit analysis of all the electroweak precision data in the third
generation quark sector (main observables are $R_b$ and $A^{FB}_b$ \cite{AFB-Rold,RSAFB})
taking into account simultaneously the mixing with KK fermions and KK gauge 
bosons, the lowest $m_{b_2}$ values would be ruled out. At this level, 
we note that in the mentioned variations of the
present minimal RS scenario, e.g. with an extended custodial symmetry or other fermion
representations, light custodians exist in general and would have different
electroweak constraints from the $b'$ here.

In conclusion, we have pointed out that in extra--dimensional models with warped
geometry and a bulk custodial symmetry, as suggested by electroweak precision
data, a new $b'_R$ quark is expected to be relatively light and will contribute
to the Higgs--gluon--gluon amplitude. It could affect in a significant way the
Higgs boson cross section in the main production channel at the LHC, $gg \to H$.
The production rates can be enhanced by a factor of four at most, compared to
the SM case. Even if the $b'$ quark is heavier than 1
TeV, i.e. beyond the reach of the LHC \cite{bLHC}, a modification of the $gg \to
H$ production rate and the $H\to gg$ decay branching ratio at the level of a few
10\% is possible.  As the KK excitations of gauge bosons and fermions are
heavy and not easy to detect at the LHC \cite{KKLHC},  the
modification of the $gg \to H$ production cross section  could be the only sign
of warped extra--dimensions. For low $b'$ masses,
the new Higgs production channel $pp \to Hb' \bar b'$ is comparable, in rate, 
to the $H t \bar t$ production.
The decays of the Higgs boson can also be affected
by the presence of the new quark but the effects 
could be probed probably only in a high--precision experiment.\\

\noindent \textbf{Acknowledgments:} This work is supported by the French  ANR
for the project {\tt PHYS@COL\&COS}. G.M. thanks E.~Kou and A.~M.~Teixeira
for useful discussions.


\begin{thebibliography}{999}
\bibitem{Mass} We do not attempt  to reproduce the precise experimental mass
values  as we neglect flavor mixing effects in the mass matrices and possible
small deviations of the Yukawa couplings from their universal values; these
approximations should not modify  our discussion, though.
\end{thebibliography}
\end{document}